# Hominin evolution was caused by introgression from Gorilla

Johan Nygren

2018

Independent researcher

Corresponding author: Johan Nygren, johanngrn@gmail.com, Independent researcher

ABSTRACT: The discovery of *Paranthropus deyiremeda* in 3.3–3.5 million year old fossil sites in Afar, together with 30% of the gorilla genome showing lineage sorting between humans and chimpanzees, and a NUMT ("nuclear mitochondrial DNA segment") on chromosome 5 that is shared by both gorillas, humans and chimpanzees, and shown to have diverged at the time of the *Pan-Homo* split rather than the *Gorilla/Pan-Homo* split, provides conclusive evidence that introgression from the gorilla lineage caused the *Pan-Homo* split, and the speciation of both the *Australopithecus* lineage and the *Paranthropus* lineage.

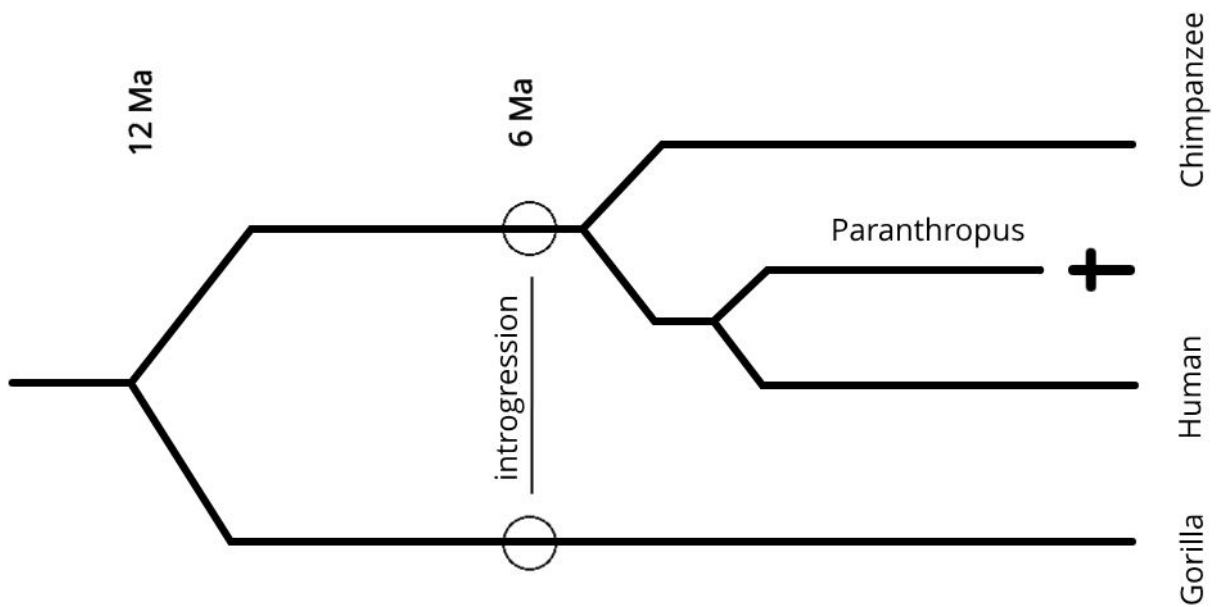

Fig. 1. Phylogenetic tree showing how introgression caused the speciation of humans. This introgression speciation model predicts an early split for *Paranthropus* and *Australopithecus*, increasingly shown in the fossil record (Haile-Selassie, 2015, 2016; Wood, 2016), and also shows that the evolution of genes that ended up in *Australopithecus*, and therefore in extant humans, as well as in *Paranthropus*, can and should be traced along the gorilla lineage as well.

# Author Summary

The exact nature of the original speciation event leading to the origin of the human and chimpanzee lineages is unknown. With advances in genome sequencing, and two decades of data on *Homo*, *Pan*, and *Gorilla*, there is now conclusive evidence that introgression from *Gorilla* caused the human-chimpanzee split, and that *Homo* and *Pan* diverged through lineage sorting with 15% of the introgressed genes ending up in *Homo* and another 15% in *Pan*. The definitive proof comes in the form of a NUMT ("nuclear mitochondrial DNA segment") on chromosome 5, tentatively called "ps5", that was transferred from *Gorilla* to *Homo* and *Pan* at the time of the *Pan-Homo* split. The reason "ps5" provides definitive proof is that mitochondrial pseudogenes like ps5 have the property that they can be compared with the mitochondrial genome, making it possible to compare the time when ps5 diverged between *Gorilla*, *Homo* and *Pan* to when the mtDNA of *Gorilla*, *Homo* and *Pan* diverged.

# Introduction

## The Gorilla lineage as the "missing link"

> *"During many years I collected notes on the origin or descent of man, without any intention of publishing on the subject, but rather with the determination not to publish, as I thought that I should thus only add to the prejudices against my views." - Charles Darwin, 1871*

Genome sequencing has been evolving along the law of accelerating returns (Kurzweil, 2004), the total amount of sequence data produced doubling approximately every seven months (Stephens, 2015). With the genetic revolution, phylogenetic relationships are no longer limited to morphological characters, they can instead be read like an open book. The fossil record, when combined with genomics, can reveal an evolutionary history that were unimaginable based on just morphological analyses. This thesis will explore a new chapter, that shows how hominin evolution is not a single continuous lineage, instead the hybridization of two separate lineages, separated over millions of years, whose genomes recombined into the hybrid lineages *Paranthropus* and *Australopithecus*. Curiously, that hybridization also accounts for the "missing link", the hybridization of two lineages explains the absence of a single continuous lineage.

The protagonist of the thesis is a single gene, a pseudogene on chromosome 5, tentatively called "ps5", that originates from the mitochondrial genome and belongs to a class of genes that have unique properties for tracing hybridization where it would have otherwise been impossible to read (Perna, 1996; Bensasson, 2001; Hazkani-Covo, 2010). This pseudogene alone provides definitive evidence that there was gene transfer between *Gorilla*, *Pan* and *Homo* at the time of the *Pan-Homo* split.

With clear evidence of introgression, the rest of the genetic trail of hybridization can be read with ease, standing on a strong foundation of indisputable proof.

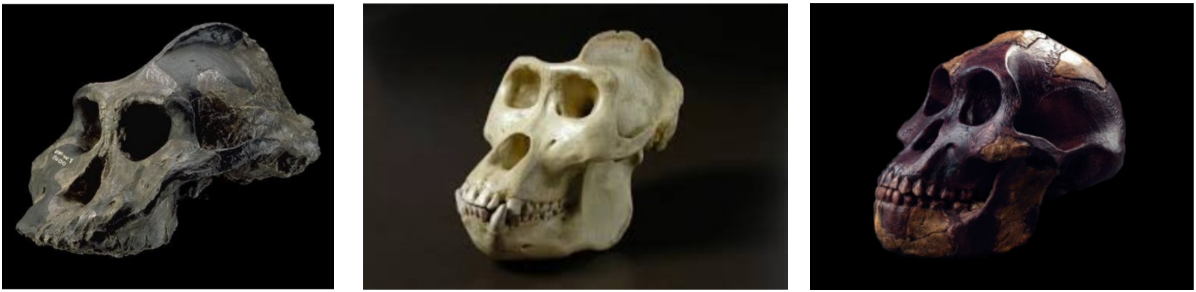

**Paranthropus aethiopicus**   **Gorilla gorilla**                **Australopithecus afarensis**
2.8-2.3 Ma                                                        3.7-2.9 Ma

Fig 2. Introgression from *Gorilla* caused the speciation of both *Australopithecus* and *Paranthropus*, and means that traits that have evolved independently in the gorilla lineage were transferred into the hybrid lineages. *Paranthropus* are often described as "gorilla-like", they have sagittal crests which suggest strong muscles of mastication, and broad, grinding herbivorous teeth, that led to the name "nutcracker man" for *Paranthropus boisei* who lived between 2.4–1.4 Ma.

# Ps5

In early screening of mitochondrial pseudogenes within the human genome, a pseudogene sequence on chromosome 5 was discovered (Li-Sucholeiki et al., 1999), which later turned out to be a large (~9kb) NUMT, tentatively called "ps5" (Popadin, 2017). With advances in genome sequencing of *Gorilla* and *Pan*, the same ~9kb pseudogene sequence was discovered at homologous chromosomal positions in both those lineages, while it was absent in *Pongo*.

The pseudogene, when compared to mitochondrial branches of *Gorilla*, *Pan* and *Homo*, is shown to have diverged between the three lineages not at the *Gorilla/Pan-Homo* split, rather at the *Pan-Homo* split (Popadin, 2017), clear evidence that there was gene transfer between the three lineages at that time.

The ps5 pseudogene shares affinities with the gorilla lineage mtDNA (Popadin, 2017) which suggests that it originated in the gorilla lineage. With the probability of a NUMT insertion being unaffected by hybridization, it is clear that the insertion happened prior to the introgression event, and that the pseudogene had been evolving in the gorilla lineage for a period of time before introgressing into *Pan* and *Homo*. (Popadin, 2017)

With high availability of genetic data for both mitochondrial DNA and the pseudogene sequence, the exact history of ps5 can be read by comparing mutations within all three lineages.

The ratio of synonymous to non-synonymous mutations is a marker to distinguish between coding and non-coding gene sequences, because non-synonymous mutations are selected against until the gene is inactivated (Tomoko, 1995). For the "stem" of the ps5 pseudogene (the mutations that have accumulated prior to its divergence into three lineages), the fraction of coding ("mitochondrial") mutations to non-coding ("pseudogenic") mutations is 3/4 (Popadin, 2017).

The mutation rate in the mitochondrial genome is significantly higher than in the nuclear genome, which means that the 25% pseudogenic mutations have needed proportionally longer time to accumulate. With the estimate of 10x higher mutation rates in mtDNA (Brown, 1979), and 3x more "mitochondrial" mutations, it took 3.3x longer to accumulate the "pseudogenic" mutations, giving a rough estimate of the insertion happening at 1.8 Myr after the *Gorilla/Pan-Homo* split, 4.2 Myr before the introgression event that led to the *Pan-Homo* split.

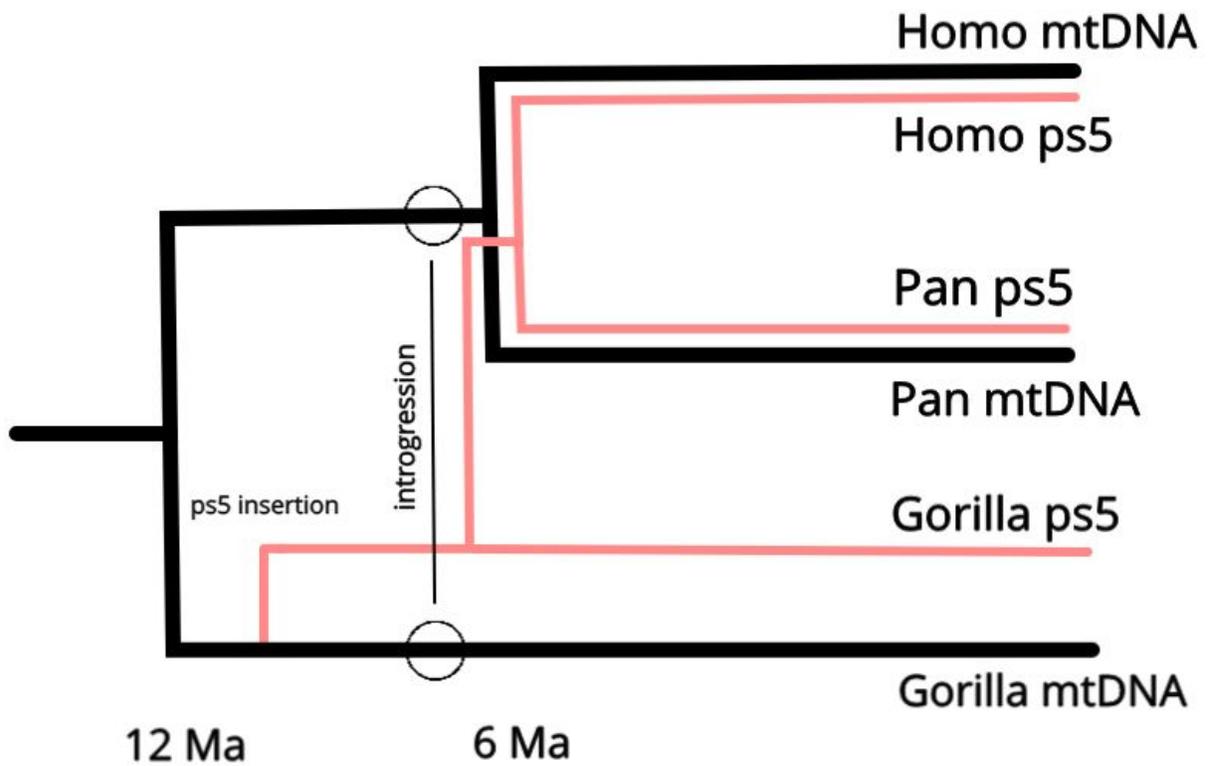

Fig. 3. Joint phylogenetic tree of hominine mtDNA and the ps5 pseudogene of mtDNA. Black and pink lines depict the mitochondrial and the pseudogene lineages respectively, diverging from their mitochondrial common ancestor. The insertion of mtDNA fragments into the nuclear genome of *Gorilla* can be roughly estimated to 1.8 Myr after the *Gorilla*/*Pan-Homo* split, and the transfer to *Pan* and *Homo* to the human-chimpanzee split, along with 30% of the Gorilla genome.

# Insights into hominin evolution from the Gorilla Genome Project

The Gorilla Genome Project was the first complete genome of *Gorilla*, from a female western lowland gorilla, and it revealed a closer relationship between humans and gorilla than what morphological analyses had shown: in 30% of the genome, gorilla is closer to human or chimpanzee than the latter are to each other. At the time interpreted as incomplete lineage sorting (Scally, 2012), the ps5 NUMT as definitive evidence of gene transfer between *Gorilla*, *Pan* and *Homo* around the time of the *Pan-Homo* split (Popadin, 2017), shows that the lineage sorting is more parsimonious as a result of introgression.

> **Introgression**
>
> Introgression is the transfer of genetic information from one species into the gene pool of another by repeated backcrossing of an interspecific hybrid with one of its parent species.

Introgression may lead to speciation, in which the new hybrid lineages become reproductively isolated from parental populations (Baack, 2007), and since *Pan* and *Homo* have diverged through lineage sorting, with 15% of the introgressed genes ending up in *Pan* and another 15% in *Homo*, it is reasonable to conclude that the introgression caused the *Pan-Homo* split (Fig. 1), and therefore that it occurred at the time of the *Pan-Homo* split, around 6 million years ago.

# Paranthropus, a companion to Australopithecus

With conclusive evidence that introgression from *Gorilla* caused the *Pan-Homo* split, it can also be seen that *Paranthropus* and *Australopithecus*, as two separate lineages, both speciated as a result of introgression from the *Gorilla* lineage (Fig. 2). The lineage sorting seen in *Pan* and *Homo* (Scally, 2012) can be predicted for *Paranthropus* as well, with the gorilla-like features, such as strong muscles of mastication, being a result of lineage sorting from the introgression of *Gorilla* (Fig. 1), conserved because the browsing adaptations that are seen in *Gorilla* were co-opted for grazing (Cerling, 2017), in convergent evolution with other species in the Afar region, such as *Eurygnathohippus* (Melcher, 2013) and *Theropithecus* (Levin, 2015), both grass-eating species descended from browsers.

## The Burtele foot (BRT-VP-2/73) and Au. deyiremeda, a Paranthropus?

The discovery of 3.2-3.5 million year old hominin fossils that show divergent evolution from *Au. afarensis* from the same time period (Haile-Selassie, 2012, 2015), featuring an abductable great toe (Fig. 4) instead of the human-like hallux of *Au. afarensis,* a human-like transverse arch that stiffens the foot (Haile-Selassie, 2012), instead of the transitional arch of *Au. afarensis* that is in-between *Homo* and *Pan*, and jaws and teeth that shares characteristics with *Paranthropus* and *Homo* (Haile-Selassie, 2015) suggested the classification of a new species *Australopithecus deyiremeda*, meaning "close relative" in the local Afar language.

The definitive proof that introgression caused the speciation both *Paranthropus* and *Australopithecus* shows that *Au. deyiremeda* is better classified as a *Paranthropus*, *P. deyiremeda*, and that an early split between *Paranthropus* and *Australopithecus*, via the same lineage sorting that is seen in *Pan* and *Homo*, is the reason there were two separate lineages of hominins during the Pliocene (Haile-Selassie, 2015, 2016; Wood, 2016), clearly distinguishable by their locomotor adaptation and diet.

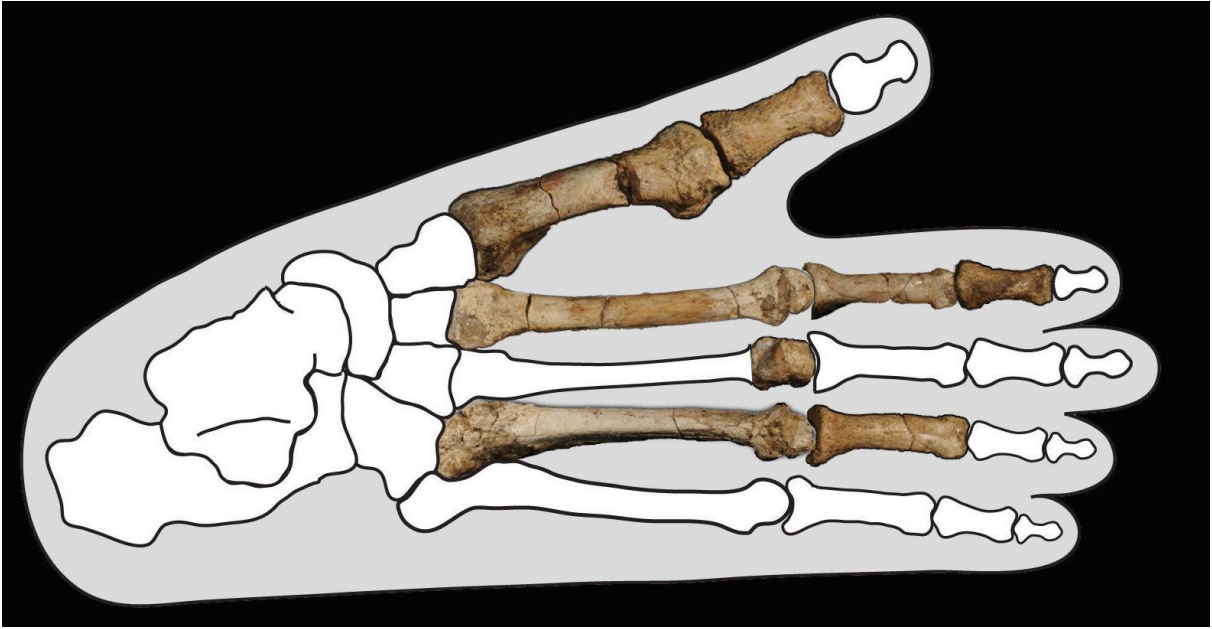

Fig 4. The Burtele foot, BRT-VP-2/73, found in 2009 (Haile-Selassie, 2015) in Burtele at Woranso-Mille, Afar, tentatively assigned *Au. deyiremeda* (Haile-Selassie, 2015), contemporaneous with *Au. afarensis*, shows distinct locomotor adaptation as it retains a grasping hallux, in contrast to the human-like adducted hallux that had developed in *Australopithecus afarensis*. The conclusive evidence that hominin evolution was caused by introgression from Gorilla suggests that *Au. deyiremeda* is better classified as *Paranthropus deyiremeda*. With a revised taxonomic classification, building on a combination of genomic data and fossil records, it can be predicted that *Paranthropus* and *Australopithecus*, like *Pan* and *Homo*, diverged through lineage sorting as the two lineages co-opted genes from the Gorilla lineage to adapt for separate niches.

# Results

## Pan-Homo split via Gorilla introgression

The lineage sorting of 30% of the gorilla genome that is seen in humans and chimpanzees (Scally, 2012) is a result of introgression, an event that caused the speciation of *Pan* and *Homo* (Fig. 1), and the two lineages diverged through lineage sorting with 15% of the introgressed genes ending up in *Pan* and another 15% in *Homo*.

## Paranthropus and Australopithecus were hybrid lineages

Traits within *Paranthropus* that resemble *Gorilla*, such as the sagittal crest (Fig. 5), are more parsimonious as a result of the introgression event rather than convergent evolution, and lineage sorting similar to the 30% of the Gorilla genome that displays lineage sorting with *Pan* and *Homo* (Scally, 2012), which supports the hypothesis of *Paranthropus* as a lineage that also speciated from the introgression (Fig. 1).

## The taxonomic classification of Paranthropus deyiremeda

The combination of data from genome sequencing with the fossil record provides an insight into how *Paranthropus* and *Australopithecus* are related, and shows that both lineages speciated as a result of introgression from *Gorilla*, and provides a foundation for the taxonomic classification of *Paranthropus deyiremeda.*

The foot stiffness in *Au. deyiremeda* (Haile-Selassie, 2012) is not a preserved character, it is a derived character that is absent in the *Au. afarensis* lineage as well as in *Pan* and *Gorilla*, and that exists together with an abducted great toe, and is contemporary with an adducted (human-like) hallux as a derived feature in *Au. afarensis* (Haile-Selassie, 2012), substantial adaptive differences that had accumulated over significant time spans of divergent evolution, indisputable data for that *Au. deyiremeda* is a separate lineage that had adapted for a separate niche, which is also what justified its original classification as a "close relative" (Haile-Selassie, 2015). The denthognathic features that are similar to *Paranthropus* (Haile-Selassie, 2015) suggest similar dietary adaptations, and within the hypothesis of introgression as a cause of speciation, the most parsimonious explanation is lineage sorting from the introgression event, with adaptations for browsing such as large muscles of mastication that were co-opted for grazing. (Cerling, 2017)

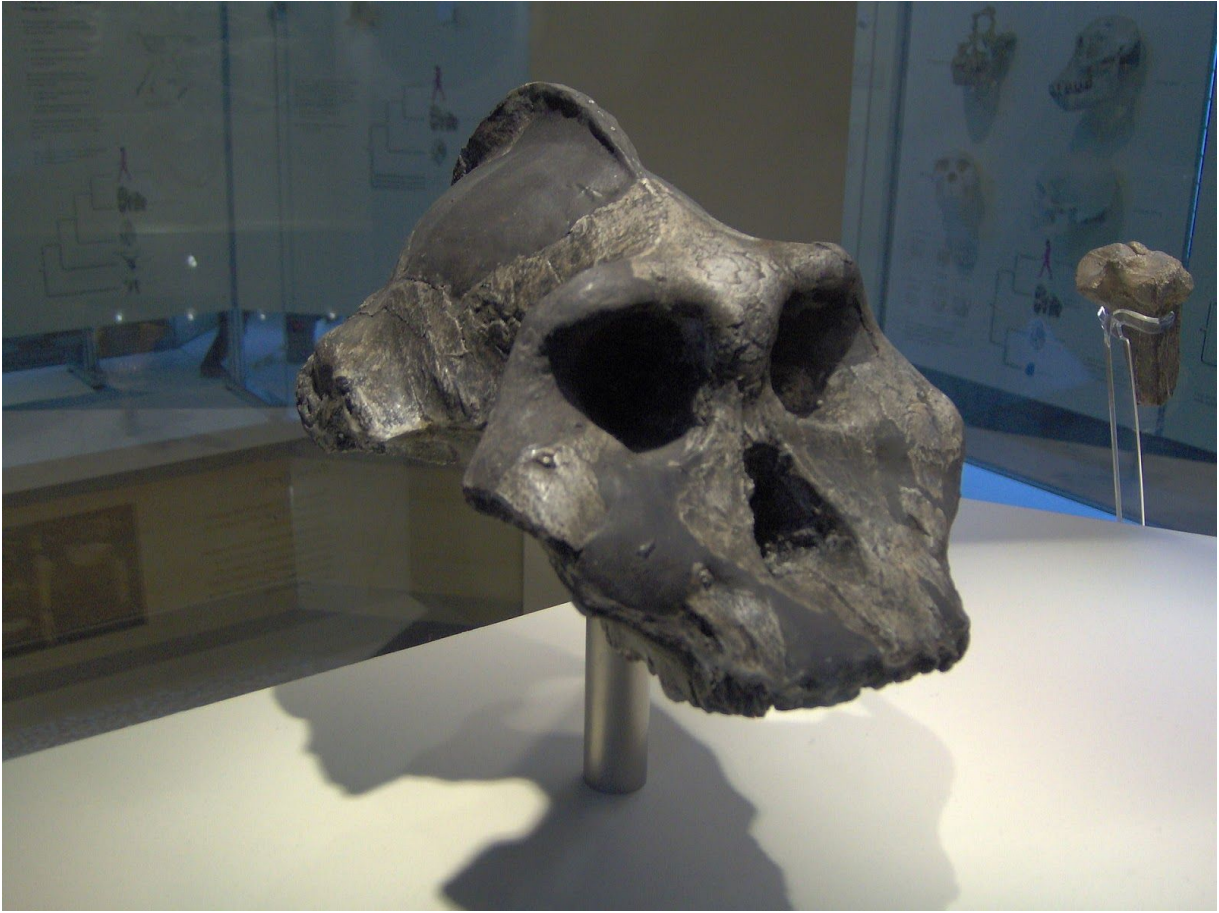

Fig. 5. *Paranthropus aethiopicus*, 2.8-2.3 Ma, with gorilla-like sagittal cranial crests as an attachment for strong muscles of mastication, a dietary adaptation. The genetic proof of an introgression event at the time of the *Pan-Homo* spit shows that the most parsimonious origin for those features within *Paranthropus* was lineage sorting from the introgression event, originating in *Gorilla*, rather than convergent evolution. Image from the public domain (CC BY-SA 3.0).

# Conclusion

The speciation of *Hominini* was caused by introgression of Gorilla, and *Pan*, *Australopithecus* and *Paranthropus* diverged as a result of lineage sorting (Fig. 1), each branch adapting for a separate ecological niche. With a strong foundation built on the genome revolution, the fossil record reveals a clear picture of two separate lineages of hybrids, *Australopithecus* and *Paranthropus*, that co-exist throughout the Pliocene and Pleistocene, and that retained separate traits from the hybridization event. The fossil evidence of *Paranthropus deyiremeda* shows that by the mid-Pliocene, the two lineages had developed separate derived features, both in foot morphology and mastication, that are later on both found within *Homo*, in other words, *Homo* has integrated traits that were separated in the ancestral *Paranthropus* and *Australopithecus* lineages.

The indisputable evidence that introgression from *Gorilla* caused the speciation of *Pan* and *Homo* is made possible by the genome revolution, centered around the mitochondrial pseudogene "ps5", and it provides a map, a reference frame, that makes it possible to read the world in ways that were previously out of sight, and can provide an important reference for continued research into hominin evolution. What remains to be understood is what environmental and ecological factors triggered the hybridization.

# Materials & Methods

## Lineage sorting and the ps5 NUMT

Phylogenetic relationships can be read from genome comparison. Mitochondrial pseudogenes within the nuclear genome, that originate from mtDNA, provide an ideal marker for tracing hybridization events over large evolutionary time scales, and the ps5 NUMT in *Gorilla*, *Homo* and *Pan* has preserved a record of an event in hominin evolution that, when combined with the fossil record as well as genome analyses as a whole, shows a clear trail of introgression from *Gorilla* at the *Pan-Homo* split, and that this hybridization was what caused the the speciation of hominins.

The Ps5 homologs in *Gorilla*, *Pan* and *Homo*, when compared to their mitochondrial genomes, shows that it formed from mtDNA at a point after the *Gorilla/Pan-Homo* split, and that it originated in the *Gorilla* lineage, with a rough estimate of insertion into the nuclear genome 1.8 Myr after the *Gorilla/Pan-Homo* split, and that it evolved within the nuclear genome of Gorilla over 3.3x the time period it accumulated "mitochondrial" mutations, to then be transferred to the common ancestor of *Pan* and *Homo* during the hybridization event, where ps5 is a small but important record of that event.

With the exponential growth rate of genome sequencing, the amount of sequence data produced doubling approximately every seven months (Stephens, 2015), there is full genome sequences for both *Homo*, *Pan* and *Gorilla* (Venter, 2001; Waterson; 2005; Scally, 2012), and the comparison of all three lineages showed, quite unexpectedly, that in 30% of the *Gorilla* genome, gorilla is closer to human or chimpanzee than the latter are to each other. In other words, there is a genomic record of lineage sorting between *Pan* and *Homo* for 30% of the *Gorilla* genome, with 15% ending up in *Pan* and another 15% in *Homo*.

Knowing that there was gene transfer at the time of the *Pan-Homo* split, the lineage sorting is most parsimonious as a result of introgression of *Gorilla*, in a hybridization event that also transferred the ps5 NUMT from Gorilla to the common ancestor of *Pan* and *Homo*, and that led to the *Pan-Homo* split as the two lineages diverged through lineage sorting of the introgressed genes (Fig. 1).

## The fossil record combined with genomics

With ps5 as definitive proof of gene transfer at the *Pan-Homo* split, and *Pan* and *Homo* showing lineage sorting from *Gorilla* providing clear evidence of introgression, a combination of that data from genome sequencing together with the fossil record can provide an insight into how *Paranthropus* and *Australopithecus* are related.

The lineage sorting seen in *Pan* and *Homo* (Scally, 2012) can be predicted for *Paranthropus* as well, with gorilla-like features, such as a sagittal crest from strong muscles of mastication, a result of lineage sorting from the introgression of *Gorilla* (Fig. 1), conserved because the browsing adaptations seen in *Gorilla* were co-opted for grazing (Cerling, 2017).

Through the combination of genomics and the fossil record, a foundation for the taxonomic classification of *Paranthropus deyiremeda* can be constructed, with clear evidence of divergent morphological features in *P. deyiremeda* and *Au. afarensis*, which fits perfectly with lineage sorting between the two hybrid lineages.

The taxonomic classification of *P. deyiremeda* extends the fossil record of the *Paranthropus* lineage backwards in time to the mid-Pliocene, 3.5 Mya, and shows a clear record of that by the mid-Pliocene, the hybrid lineages *Australopithecus* and *Paranthropus* had adapted to separate niches, each lineage conserving its own set of traits from their two parental lineages.

# Acknowledgements

Johan Nygren is an independent researcher and the author of the thesis and all content in it.

# Figure Legends

**Fig. 1. Phylogenetic tree showing how Gorilla introgression caused the human-chimpanzee split**

Phylogenetic tree showing how introgression caused the speciation of humans. This introgression speciation model predicts an early split for *Paranthropus* and *Australopithecus*, increasingly shown in the fossil record (Haile-Selassie, 2015, 2016; Wood, 2016), and also shows that the evolution of genes that ended up in *Australopithecus*, and therefore in extant humans, as well as in *Paranthropus*, can and should be traced along the gorilla lineage as well.

**Fig. 2. Morphological traits in Gorilla and the hybrid lineages Paranthropus and Australopithecus**

Introgression from *Gorilla* caused the speciation of both *Australopithecus* and *Paranthropus*, and means that traits that have evolved independently in the gorilla lineage were transferred into the hybrid lineages. *Paranthropus* are often described as "gorilla-like", they have sagittal crests which suggest strong muscles of mastication, and broad, grinding herbivorous teeth, that led to the name "nutcracker man" for *Paranthropus boisei* who lived between 2.4–1.4 Ma.

**Fig. 3. Phylogenetic tree with hominine mtDNA and ps5 homologs**

Joint phylogenetic tree of hominine mtDNA and the ps5 pseudogene of mtDNA. Black and pink lines depict the mitochondrial and the pseudogene lineages respectively, diverging from their mitochondrial common ancestor. The insertion of mtDNA fragments into the nuclear genome of *Gorilla* can be roughly estimated to 1.8 Myr after the *Gorilla*/*Pan-Homo* split, and the transfer to *Pan* and *Homo* to the human-chimpanzee split, along with 30% of the Gorilla genome.

**Fig. 4. Fossil record for a taxonomic classification of Paranthropus *deyiremeda***

The Burtele foot, BRT-VP-2/73, found in 2009 (Haile-Selassie, 2015) in Burtele at Woranso-Mille, Afar, tentatively assigned *Au. deyiremeda* (Haile-Selassie, 2015), contemporaneous with *Au. afarensis*, shows distinct locomotor adaptation as it retains a grasping hallux, in contrast to the human-like adducted hallux that had developed in *Australopithecus afarensis*. The conclusive evidence that hominin evolution was caused by introgression from Gorilla suggests that *Au. deyiremeda* is better classified as *Paranthropus deyiremeda*. With a revised taxonomic classification, building on a combination of genomic data and fossil records, it can be predicted that *Paranthropus* and *Australopithecus*, like *Pan* and *Homo*, diverged through lineage sorting as the two lineages co-opted genes from the Gorilla lineage to adapt for separate niches.

**Fig. 5. Gorilla-like traits in the Paranthropus lineage a result of lineage sorting from Gorilla**

*Paranthropus aethiopicus*, 2.8-2.3 Ma, with gorilla-like sagittal cranial crests as an attachment for strong muscles of mastication, a dietary adaptation. The genetic proof of an introgression event at the time of the *Pan-Homo* spit shows that the most parsimonious origin for those features within *Paranthropus* was lineage sorting from the introgression event, originating in *Gorilla*, rather than convergent evolution. Image from the public domain (CC BY-SA 3.0).